# On the low-temperature ordering of the 3D antiferromagnetic three-state Potts model

Miroslav Kolesik* and Masuo Suzuki
Department of Physics, University of Tokyo,
Bunkyo-ku, Tokyo 113, Japan

April 22, 1995

**Abstract**

The antiferromagnetic three-state Potts model on the simple-cubic lattice is studied using Monte Carlo simulations. The ordering in a medium temperature range below the critical point is investigated in detail. Two different regimes have been observed: The so-called broken sublattice-symmetry phase dominates at sufficiently low temperatures, while the phase just below the critical point is characterized by an effectively continuous order parameter and by a fully restored rotational symmetry. However, the later phase is not the permutationally sublattice symmetric phase recently predicted by the cluster variation method.

# 1 Introduction

The properties of the three-dimensional antiferromagnetic three-state Potts model have been investigated intensively over fifteen years, but complete understanding of its low-temperature behavior is still missing. There have been disputes about its universality class as well as about the type of the low-temperature ordering. Let us review the most important works briefly. In 1980, Berker and Kadanoff [1] suggested that the low-temperature phase of the present model shows algebraically decaying correlation. Banavar et al. [2, 3] used the Monte Carlo simulation to study the ordering in various Potts models. They observed the so-called broken-sublattice-symmetry (BSS) phase, and conjectured that the model under discussion belongs to the universality class of the XY model. Ono [4] concluded that there is no spontaneous magnetization at low temperatures, but a "Kosterlitz-Thouless phase" exists below the critical point. On the other hand, Ueno[5]

---

*Permanent address: Institute of Physics, SAS, Dúbravská cesta 9, Bratislava 842 28, Slovakia

observed a nonzero magnetization and suggested that the model belongs to a new universality class. Wang et al. [6, 7] obtained critical exponents quite close to those of the XY model, and they also found a nonvanishing magnetization. Interestingly, they observed a divergent susceptibility even below the critical point (at $T \sim 0.68 T_c$), indicating that the ordered phase is characterized by algebraically decaying correlations. In 1992, Okabe and Kikuchi [8, 9] measured the probability distribution of the magnetization closely below the critical point, and they interpreted their results as an evidence of the BSS phase. Recently, Rosengren and Lapinskas investigated the Blume-Emery-Griffiths model using the cluster variation method [10], and they arrived at very interesting implications for the antiferromagnetic Potts model. They argued that the BSS phase exists only at low temperatures, while there is a new, so-called permutationally-symmetric sublattices (PSS) phase in a narrow region just below the critical point. They found two new phase transitions between the PSS and BSS phases together with an extremely narrow ferrimagnetic phase between the transitions. Ueno later argued [11] that the low-temperature phase should be characterized as an incompletely ordered phase, and that it is not compatible with the XY universality class. However, in an extensive Monte Carlo study, Gottlob and Hasenbusch [12, 13] obtained very accurate estimates of critical exponents indicating the XY universality class. Gottlob and Hasenbusch also studied the probability distribution of magnetization, and they found that it becomes rotationally symmetric when the system approaches its critical point. Their results were corroborated also by our recent calculations [14] based on the coherent-anomaly method (see e.g. review in Ref. [15]).

While the issue of the universality class seems to be resolved quite convincingly by Gottlob and Hasenbusch, there is still a surprising amount of disagreement about the type of the ordered phase. The motivation of the present work is to contribute to the solution of this interesting problem. Thus, we aim at the revealing the nature of the ordering in the region in which Wang at al. observed the divergent susceptibility, and in which the new phase transitions were predicted by Rosengren and Lapinskas. We try to answer the following questions: Is there a region with the PSS phase? If there is not, then does the BSS phase span the whole temperature region below the critical point?

The layout of the remainder of the present paper is as follows. In order to fix the notation, we describe the model and its order parameters in the next section. We also give the characteristics of various phases which we expected to see in our simulations. Monte Carlo simulations are described in Section 3, and the results are discussed in Section 4. Finally, we give some remarks concerning possible alternative interpretations in the Conclusion.

## 2 Order parameters

The Hamiltonian of the Potts model is written as

$$H = J \sum_{<i,j>} \delta(s_i, s_j) \qquad (1)$$

where the summation runs over all nearest-neighbor pairs on the cubic lattice, and the spin variables $s_i$ take on three different values, say $\{1, 2, 3\}$. In what follows, we put the interaction $J$ equal to unity, and $K$ stands for the inverse temperature.

Let us denote by $c_i^a$ and $c_i^b$ the concentrations of those spins located on the sublattice $a$ and $b$, respectively, which are in the state $i$. The concentrations $c_i^{a,b}$ are constrained by

$$\sum_{i=1}^{3} c_i^a = \sum_{i=1}^{3} c_i^b = 1 , \qquad (2)$$

and we use the following parameterization which takes into account the three-fold symmetry of the model

$$\begin{aligned} c_i^a &= 1/3 + 2/3 \ r_a \cos(\phi_a - \psi_i) \\ c_i^b &= 1/3 + 2/3 \ r_b \cos(\phi_b - \psi_i) \end{aligned} \quad \text{with} \quad \psi_{1,2,3} = 0, +2\pi/3, -2\pi/3 . \qquad (3)$$

Each couple of $\{r_a, \phi_a\}$ and $\{r_b, \phi_b\}$ denote the polar coordinates of each vector $\vec{r}_a$ and $\vec{r}_b$ in the plane. We call them sublattice magnetizations or order parameters. Their allowed values fall within the equilateral triangle which is centered at the origin and has one of its vertices at the location $\{1, 0\}$ (see Fig. 1). We also use the magnetization, which we define as the sublattice difference

$$\vec{m} = \vec{r}_a - \vec{r}_b , \quad m = |\vec{m}| \qquad (4)$$

and denote by $\phi_{\mathrm{mag}}$ the angle describing its direction.

Let us describe various phases in terms of the above order parameters. Naturally, we have vanishing sublattice magnetizations $r_a$, $r_b$ at sufficiently high temperatures. Thus, the center of the allowed triangle represents the disordered phase..

Furthermore, we expect to observe the broken sublattice symmetry (BSS) phase at very low temperatures. In such a phase, one of the sublattices is populated predominantly by one of the three Potts states, while the second sublattice is populated mainly by the remaining two states. There are six different BSS phases related by permutations of the three spin states. The angles $\phi_{a,b}$ and the magnetization angle $\phi_{\mathrm{mag}}$ can have (or should be close to, on finite lattices) only "nice" values, namely $k\pi/3$ with $k = 0, 1 \ldots 5$, and the radii $r_a$ and $r_b$ are not equal:

$$r_a \neq r_b , \quad \phi_a = k\pi/3 , \quad \phi_b = \phi_a \pm \pi \quad \text{and} \quad \phi_{\mathrm{mag}} = k\pi/3 \quad \text{with} \quad k = 0, 1 \ldots 5 . \qquad (5)$$

The locations corresponding to the states of the two sublattices are near the vertices of the allowed triangle, and in the vicinity of the middles of its sides, respectively (see Fig. 1).

Let us characterize the permutationally-symmetric sublattices phase (PSS) proposed by Rosengren and Lapinskas [10]. In this phase, the concentrations of the spin states on one sublattice are the same as the concentrations on the second sublattice, but the maximal and minimal concentrations are exchanged. There are again six PSS phases, but this time there are no special values for the angles $\phi_{a,b}$; the angle between $\vec{r}_a$ and $\vec{r}_b$ is somewhat less then $\pi$. On the other hand, the radii $r_a$ and $r_b$ are equal, and the magnetization angle $\phi_{\mathrm{mag}}$ takes on special values:

$$r_a = r_b , \quad \phi_a \sim (\neq) \ \phi_b \pm \pi \quad \text{and} \quad \phi_{\mathrm{mag}} = (k/3 + 1/2)\pi \quad \text{with} \quad k = 0, 1 \ldots 5 \qquad (6)$$

# 3 Monte Carlo simulations

Note that if we want to distinguish between the BSS and PSS phases, then the magnetization $\vec{m}$ or even its absolute value $m$ are not very useful quantities, because they are very similar in the both phases. Consequently, we have to concentrate on the difference $r_a - r_b$ which is equal to zero in the PSS phase while it remains finite in the BSS phase. We also want to see what are "allowed" values for the angles $\phi_a$, $\phi_b$ and $\phi_{\mathrm{mag}}$. Our strategy is to sample the probability distributions for interesting quantities, and to investigate the finite size effects in order to determine the type of ordering.

Because the interplay between the two sublattices is crucial in the present model, we have carried out the summation over the spin-states on one of them. Then, we simulated the resulting model with multispin interactions. The rationale behind this choice of the simulation algorithm was twofold. Firstly, it turned out in the simulations by Wang et al. [7] that the cluster-algorithm performance was much worse in the ordered phase than right at the critical point. Secondly, we have observed in our previous multilattice microcanonical simulations that the inter-sublattice relaxation was very slow in the medium-temperature range. On the contrary, the sublattices are always in a "mutual equilibrium" within the present approach.

We have simulated cubic lattices with the linear sizes $L = 8, 16, 24, 32, 40, 48$ and $64$ at several inverse temperatures within the relevant region, namely at $K = 0.9, 1.0, 1.1, 1.2, 1.3, 1.4$ and $2.0$. Each run consisted of $10^5$ sweeps for thermalization and $10^6$ sweeps for measurements with data collected at every 10th and every 100th step for various probability distributions and other order parameters, respectively.

# 4 Simulation results and interpretation

## 4.1 Ordering in the $\vec{r}_{a,b}$-plane

Figure 1 shows the instantaneous sublattice magnetizations recorded during the runs at two different temperatures. The probability distribution of the sublattice magnetization is localized along a closed curve at both temperatures. But one can see that at the lower temperature, $K = 1.4$, this probability concentrates near the points characteristic for the BSS phase. At the same time, the occurrence of the PSS-like states is very unprobable. This difference becomes more an more pronounced as the lattice size increases, and one expects that the probability distribution would converge to the six delta functions located near the vertices and/or near the side centers of the allowed triangle. On the other hand, at the higher temperature, $K = 1.0$, the PSS-like configurations seem to be equally probable as the BSS-like configurations are. Note that $K = 1.0$ is already in the region in which Rosengren and Lapinskas found the PSS phase [10]. Here we see that the PSS-like configurations evidently play an important role, but they do not seem to suppress the BSS-like states.

Figure 2 presents the results of the same measurements as Fig. 1 but viewed in the $\phi_a - \phi_b$-plane. Again, we see the tendency towards the BSS-type ordering at low temperatures, while there are no clearly dominant configurations at higher temperatures.

Now, we face the question what does this observation mean. Is it merely an extremely pronounced size effect, or are there two different ordered phases? In order to see the low-temperature behavior in a more quantitative way, we have measured several histograms which we describe below.

## 4.2 Probability distribution of the $|r_a - r_b|$

We have measured the probability distribution of the radii difference $|r_a - r_b|$ at several temperatures in an attempt to observe the PSS phase. Let us recall that this probability should be accumulated around $|r_a - r_b| = 0$ in the PSS phase (see (6)). In the BSS phase, one should observe a peak centered at a finite, temperature dependent value of the $|r_a - r_b|$ (see (5)). One can see a clear tendency towards the BSS-type ordering at low temperatures in Fig. 3, but in the medium temperature region $K \leq 1.0$, the probability of the PSS-like configurations with $|r_a - r_b| = 0$ does not seem to converge to zero as the lattice size increases. At the same time, we still see a strong peak corresponding to BSS-like configurations. Such a behavior is not compatible with the PSS phase proposed by Rosengren and Lapinskas, but it shows that something interesting *is* going on.

## 4.3 Probability distribution of the $|\phi_a - \phi_b|$

Clearly, we need some quantity which is more sensitive to the finite-size effects to look closely at the crossover region between the two regimes. We want to find out whether the probability distribution goes to some delta function(s) in the thermodynamic limit, or whether it remains accumulated along the whole closed curve in the order-parameter space even for $L \to \infty$. This is why we have measured the distribution of the $||\phi_a - \phi_b| - \pi|$. If the system tends to the BSS phase in the thermodynamic limit, then this distribution should converge to a delta function located at zero (see (5)). On the other hand, if the probability of the PSS-like configurations remains finite for arbitrarily large lattices, then we should observe a growing peak localized at the "maximal" (again temperature dependent) value of the $||\phi_a - \phi_b| - \pi|$ (see (6)). Indeed, we see pronounced size effects in Fig. 4. The PSS-like peak is strongly suppressed on larger lattices at low-temperatures. However, it seems to survive for $K \leq 1.0$. This suggests that we have really to do with two different regimes. Let us look at the finite-size effects more closely: In Fig. 5 we have shown the height of the PSS-like peak for various lattices and temperatures. At lower temperatures, its height first increases and then decreases as the lattice size increases. The position of the maximum is shifted to larger lattice sizes as the temperature rises. Only for $K \leq 1.0$ we have not seen the maximum, but the peak grows even on the lattice as large as $L = 64$. This behavior can be understood easily, at least qualitatively, as a result of two competing effects: The first one is the growth of the PSS-like peak caused by the decreasing width of the probability distribution in the angle-angle space. The second effect is the suppression of the PSS-like configurations which becomes discernible on larger lattices. This is why one expects the behavior of the PSS-peak height to be qualitatively described by

$$\text{height} \sim L^v \exp(-\delta L^p) \tag{7}$$

The first factor corresponds to the narrowing of the distribution width and its exponent $v$ is determined by the behavior of the corresponding susceptibility. The exponential factor in (7) describes the suppression of the PSS-like configurations. To determine the exponent $p$ we have to take into account the following important fact. The PSS-like configurations observed in our simulations are homogeneous, meaning that they are not configurations with an interface, or twisted configurations. We have observed that one can detect the PSS-like configurations even locally by looking at small sections of the lattice. This implies $p = 3$ with the parameter $\delta$ being the difference between the free energies corresponding to BSS- and PSS-like configurations. We have fitted the formula (7) with $p = 3$ to the data shown in Fig. 5, and obtained a good agreement. We must stress that the resulting values of the free energy difference $\delta$ cannot be taken too seriously, because the values of the exponent $v$ cannot be determined very accurately. Nevertheless, it turns out that the size dependence of the PSS-peak height at $T = 1.0$ is fully compatible with a pure power law, and that it increases approximately linearly with the lattice size for $K = 0.9$. This indicates that the PSS-like configurations are not suppressed any more in this temperature region.

## 4.4 Alternative order parameter

In order to get rid of the first of the two finite-size effects above mentioned, namely of the effect of the decreasing width of the order-parameter probability distribution, which is not interesting now, we have measured the order parameter related to the magnetization angle:

$$a = \langle \cos(6\phi_{\mathrm{mag}}) \rangle \quad (8)$$

Let us recall that the values of $\phi_{\mathrm{mag}}$ expected in the BSS and PSS phases are $k\pi/3$ and $(k/3 + 1/2)\pi$, respectively. Thus, the parameter $a$ equals one in the BSS phase, while it would be $-1$ in the PSS phase. More precisely, it measures the deviation from the perfect rotational symmetry of the magnetization probability distribution: In a rotationally symmetric phase one has $a = 0$. We call it the asymmetry parameter.

Let us estimate, at least roughly, the finite-size effects for this parameter. For this purpose, we introduce the restricted partition function

$$Z_L(\phi_1, \phi_2) \equiv \exp(-\beta F_L(\phi_1, \phi_2)) = \sum_{\phi_{\mathrm{mag}} \in \langle \phi_1, \phi_2 \rangle} \exp(-\beta H) \quad (9)$$

with the summation running only over the configurations which have the magnetization angle $\phi_{\mathrm{mag}}$ in the interval $\langle \phi_1, \phi_2 \rangle$. Then, the asymmetry order parameter can be formally expressed as

$$a = \langle \cos(6\phi_{\mathrm{mag}}) \rangle = \int_0^{2\pi} \cos(6\phi) \exp(-\beta F_L(\phi, \phi + d\phi)) d\phi \bigg/ \int_0^{2\pi} Z_L(\phi, \phi + d\phi) d\phi \quad (10)$$

Our simulations indicate that, in the medium temperature region, the restricted free energy $F_L(\phi, \phi + d\phi)$ becomes extremely flat as a function of the magnetization angle $\phi_{\mathrm{mag}}$. When the free energies of the BSS-like and PSS-like states are close to each other, then the variation of $F_L$ is expected to behave as $\Delta L^3 \cos(6\phi)$ as a consequence of the model symmetry. From this follows the expression in terms of the Bessel functions

$$a(L) = I_1(\Delta L^3)/I_0(\Delta L^3) \quad (11)$$

Here, the free-energy difference $\Delta$ is essentially the same quantity as $\delta$ in (7) and is supposed to exhibit only a weak size dependence.

Figure 6 shows the lattice size dependence of the asymmetry $a$ for various temperatures together with the fits obtained from (11) by adjusting $\Delta$. One can see that the agreement is reasonable. The free energy difference $\Delta$ decreases rapidly as the temperature rises towards $T \sim 1.0$; we have obtained the values $3.1 \times 10^{-4}$, $3.3 \times 10^{-5}$, $1.6 \times 10^{-5}$, $5.5 \times 10^{-6}$, $2.1 \times 10^{-6}$ for the inverse temperatures $K = 2.0, 1.4, 1.3, 1.2, 1.1$, respectively. With our data, $\Delta$ is indistinguishable from zero for $K = 1.0$ and $K = 0.9$. This is a strong indication that the region below $T \sim 1.0$ is the BSS phase, while there is another ordered phase for $T > 1.0$ which is rotationally symmetric. In the later, all the states in which the magnetization points in an arbitrary direction but has the same absolute value, are equally probable. Unfortunately our data are by far insufficient to locate the transition precisely. We can give only a rough estimate that it is located close below $T = 1.0$: This is seen also in Fig. 7, which presents the asymmetry temperature dependence for different lattices. We observe that $a$ is nearly equal to one for large lattices and at low temperatures, and drops suddenly to a value which we are not able to distinguish from zero at temperatures $T \geq 1.0$.

## 5 Conclusion

In conclusion, our Monte Carlo simulations provide a strong indication that there are two different ordered phases below the critical point in the model under investigation. The very low-temperature phase is the expected broken-sublattices symmetry (BSS) phase. In the medium temperature region $T \in (\sim 1.0, T_c = 1.2215...)$, we have observed a phase, in which the magnetization probability distribution acquires a full rotational symmetry. The transition temperature between the two regimes is probably close below $T = 1.0$.

At the present stage we also cannot rule out the scenario in which the difference between the free energies of the BSS and the PSS phases vanishes only at the critical point, and is extremely small within a whole medium temperature region. Unfortunately, very long simulations on very large lattices would be needed to discriminate between the two possibilities. The problem is that the parameter $a$, which is probably the most sensitive quantity to measure the deviations from the rotational symmetry of the magnetization distribution, is very difficult to measure accurately, because its correlation time is related to the "ergodic time" which the system needs to accomplish a full rotation in the order-parameter space. At present it is not clear to us what would be the best way to discriminate between the two scenarios and, possibly, to locate the phase transition accurately.

We would like to make few remarks concerning the ordering at low temperatures and relate our observations to previous results.

It is interesting to observe that the measured values of the order parameters are localized along definite curves (see Fig. 2). However, it is not difficult to explain their origin qualitatively. Let us consider a spin together with its six neighbors. There are $2^6$ different configurations of the neighboring spins. For each such configuration we can calculate the probability distribution for the spin at the center; it represents a point in the order parameter space, and we can observe that these points are localized along the curves found in the simulations. This tells us that the shape of the probability-distribution locus is determined mainly by the local interaction.

The second remark concerns previous studies by Wang et al. [7] in which they observed a divergent susceptibility related to the width of the distribution of the absolute value of the magnetization. They found that it diverges linearly with the lattice size at the temperature $T = 0.68T_c$. We also have measured this susceptibility. Our results confirmed the observation by Wang et al. in the sense that the susceptibility is divergent. Nevertheless, we observed that the rate of divergence is dependent on the temperature. We have found a linear increase at $1/T = 1.2$ which is very close to the above $T = 0.68T_c$, but the increase is sub- (super) linear for higher (lower) temperatures. Similarly, the behavior of the PSS-like peak in the probability distribution of the angle difference is an indirect evidence that the corresponding susceptibility diverges.

It is also worth to make the following remark concerning the results of the mean-field type methods such as the cluster variation method used by Rosengren and Lapinskas [10] or the variational series expansion approach applied in our recent study [14]. These approaches give the "right answer" in a sense, namely that the free energies of the BSS and PSS phases are very similar. However, the sign of the very small relative difference calculated by such an approximation need not necessarily correspond to reality. Therefore, we should be careful about drawing conclusions regarding the phase diagram based on mean-field type methods, especially in this peculiar region of the Potts model.

## Acknowledgments

One of us (M.K.) would like to express his gratitude to the Nishina Memorial Foundation for granting him a scholarship. The computer simulations were performed on the Hewlett-Packard workstations of the Suzuki group, Department of Physics, and on the HITAC S3800/480 of the Computer Center, University of Tokyo.

**Fig. 1.** The instantaneous sublattice magnetization recorded during the simulation runs on the lattice of the linear size $L = 48$ at two different inverse temperatures $K$. Note that the probability distribution at the lower temperature is accumulated in the vicinity of the six points corresponding to the six different BSS phases (two of them are indicated by arrows), and the occurrence of the PSS-like configurations (again, two of the locations are indicated by arrows) is strongly suppressed. On the other hand, at higher temperatures, the probability is nearly constant along the whole "curve".

Some items are shown to make the notation in the text clear: The triangle is the boundary of the allowed region for the sublattice magnetizations. Sublattice magnetizations $\vec{r}_a$ and $\vec{r}_b$ together with the resulting magnetization $\vec{m}$ are shown in a typical PSS-like configuration. The spin concentrations $\{c_i\}$ corresponding to a point within the triangle are given as its projections onto the $c_i$ axes which have their origins in the middles of sides of the triangle. The vertices of the triangle would represent the fully ordered sublattices.

Fig. 2

**Fig. 2.** The instantaneous sublattice angles $\phi_a$ and $\phi_b$ recorded as in Fig. 1. Only one half of the probability-distribution locus is shown (note that it is symmetric with respect to the reflection $\phi_a \leftrightarrow \phi_b$ as well as with respect to the translation $\phi \to \phi + 2/3\pi$). The locations of the BSS- and PSS-like configurations are shown. At lower temperatures, the PSS configurations have a very small probability to be observed, while at $K = 1.0$, the probability distribution does not vary along the curve. The shape of probability distribution locus is determined by the local interactions as is explained in Conclusion.

Fig. 3a

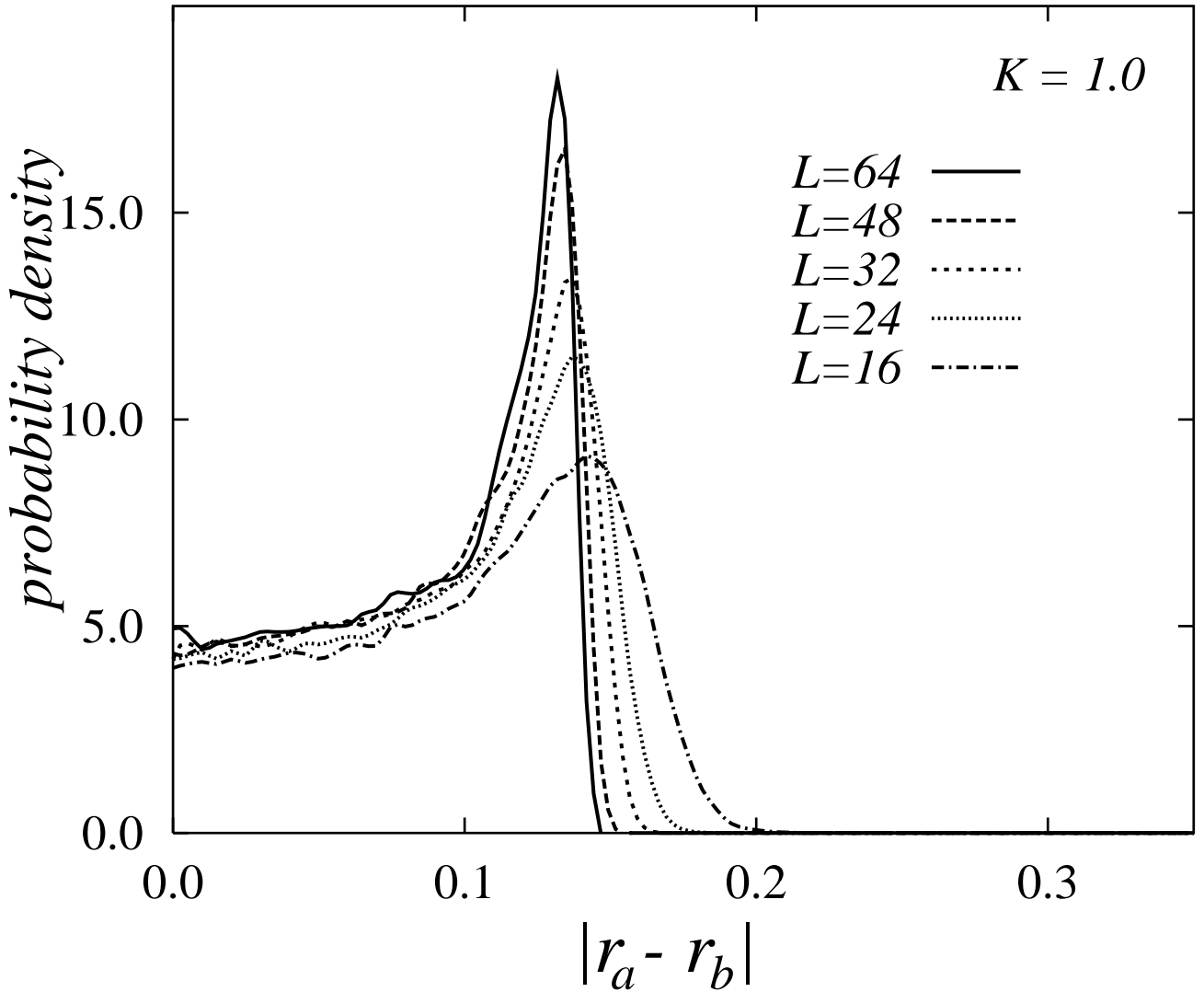

**Fig. 3.** The probability distributions for the absolute value of the difference $r_a - r_b$ measured for various lattice sizes and at different inverse temperatures: a) $K = 1.0$, b) $K = 1.2$, c) $K = 1.4$. The region around the origin corresponds to the PSS-like configurations, and the peaks at "maximal values" represent the BSS-like configurations. At the lowest temperature (c), one can see that the probability distribution tends to a single peak with the increasing lattice size $L$. It is expected to converge to a delta function located at a finite, temperature dependent value of $|r_a - r_b|$. At the medium temperature (b), the suppression of the probability at the origin is still discernible, but at the highest temperature (a), we do not observe the suppression of the PSS-like configurations any more. This suggests that in the $L \to \infty$ limit we would obtain a probability distribution which is nonzero on an interval $\langle 0, \Delta r_{\max} \rangle$ and divergent at $\Delta r_{\max}$.

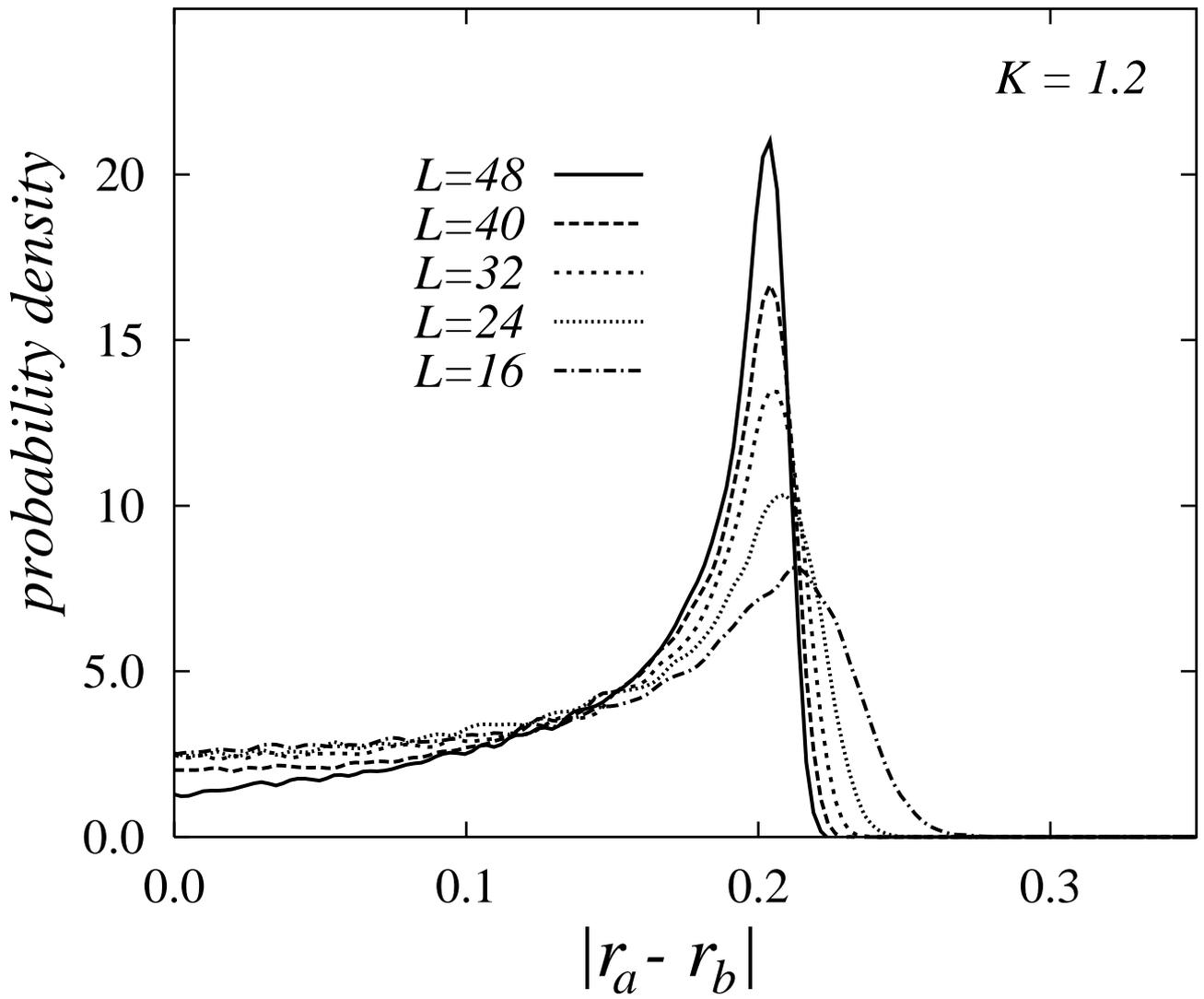

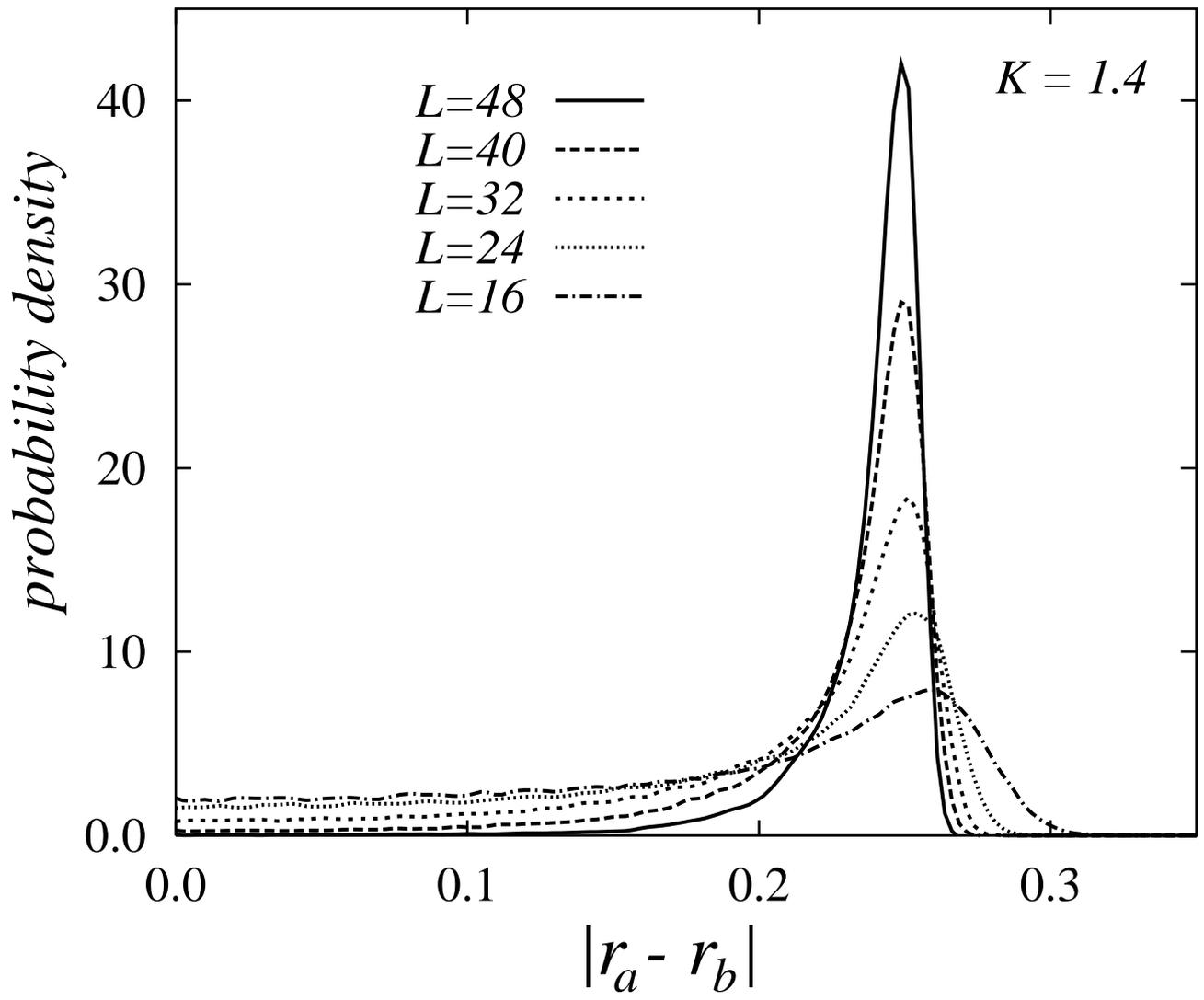

Fig. 4a

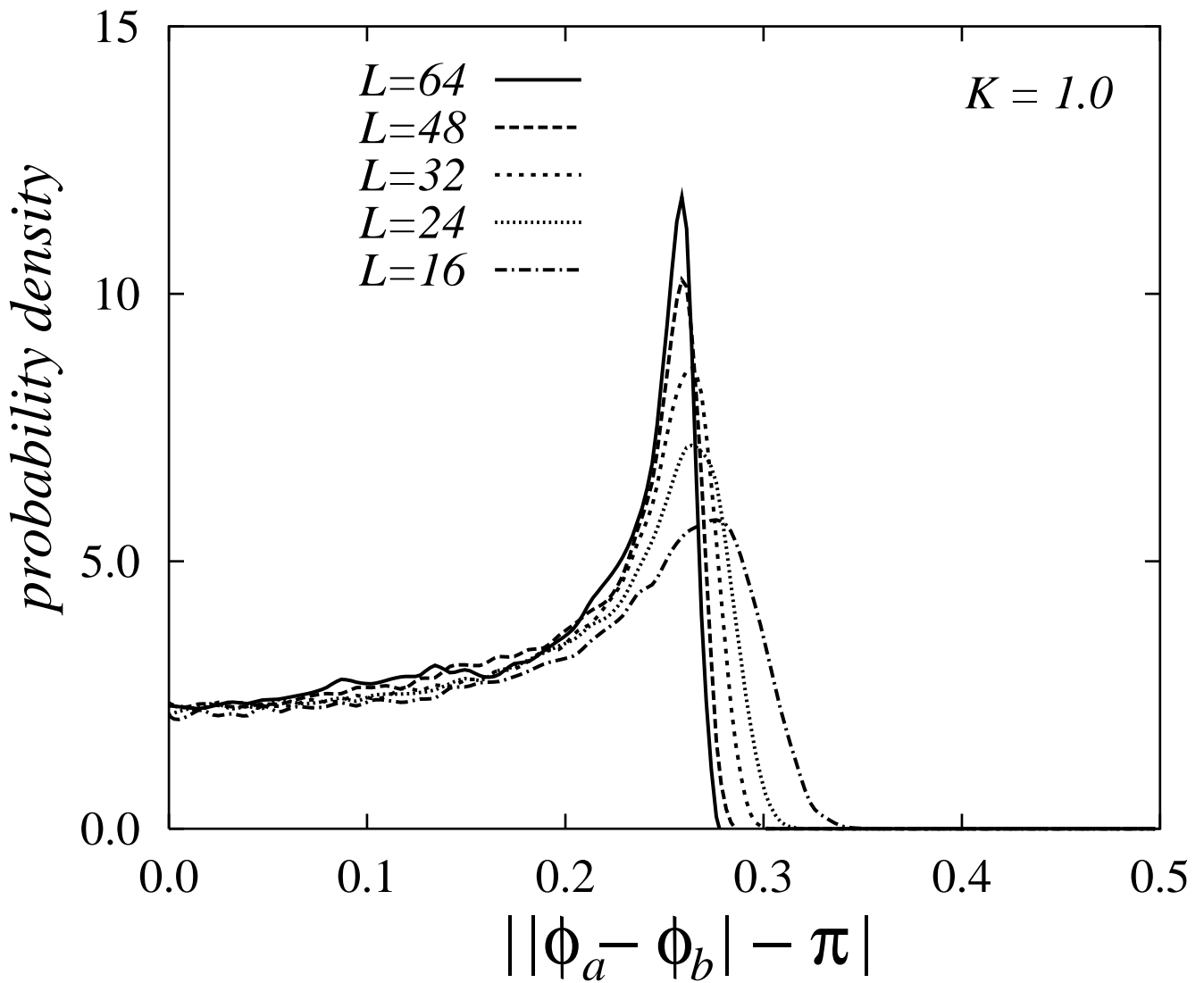

**Fig. 4.** The probability distribution for the $||\phi_a - \phi_b| - \pi|$ on various lattices and at three different temperatures: a) $K = 1.0$, b) $K = 1.2$, c) $K = 1.4$. This is a complementary picture compared to Fig. 3; The peaks at the maximal values are due to PSS-like configurations, and the BSS-configurations are seen at the origin. The PSS-like configurations die-out at lower temperatures as the lattice size increases, and the probability distribution forms a peak located at the origin; this is a pure BSS phase. At higher temperatures, both the BSS- and PSS-like configurations are relevant.

Fig. 4b

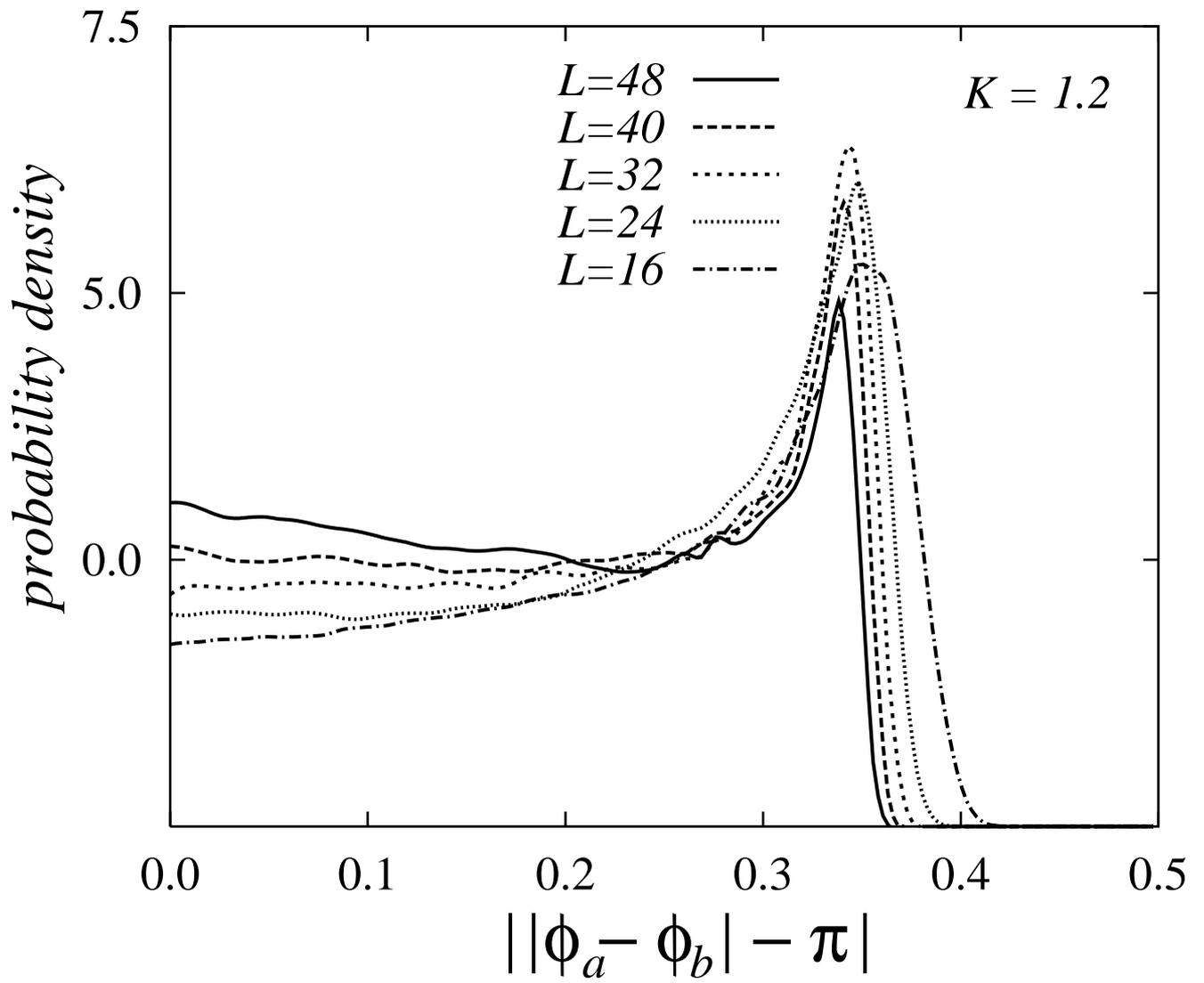

Fig. 4c

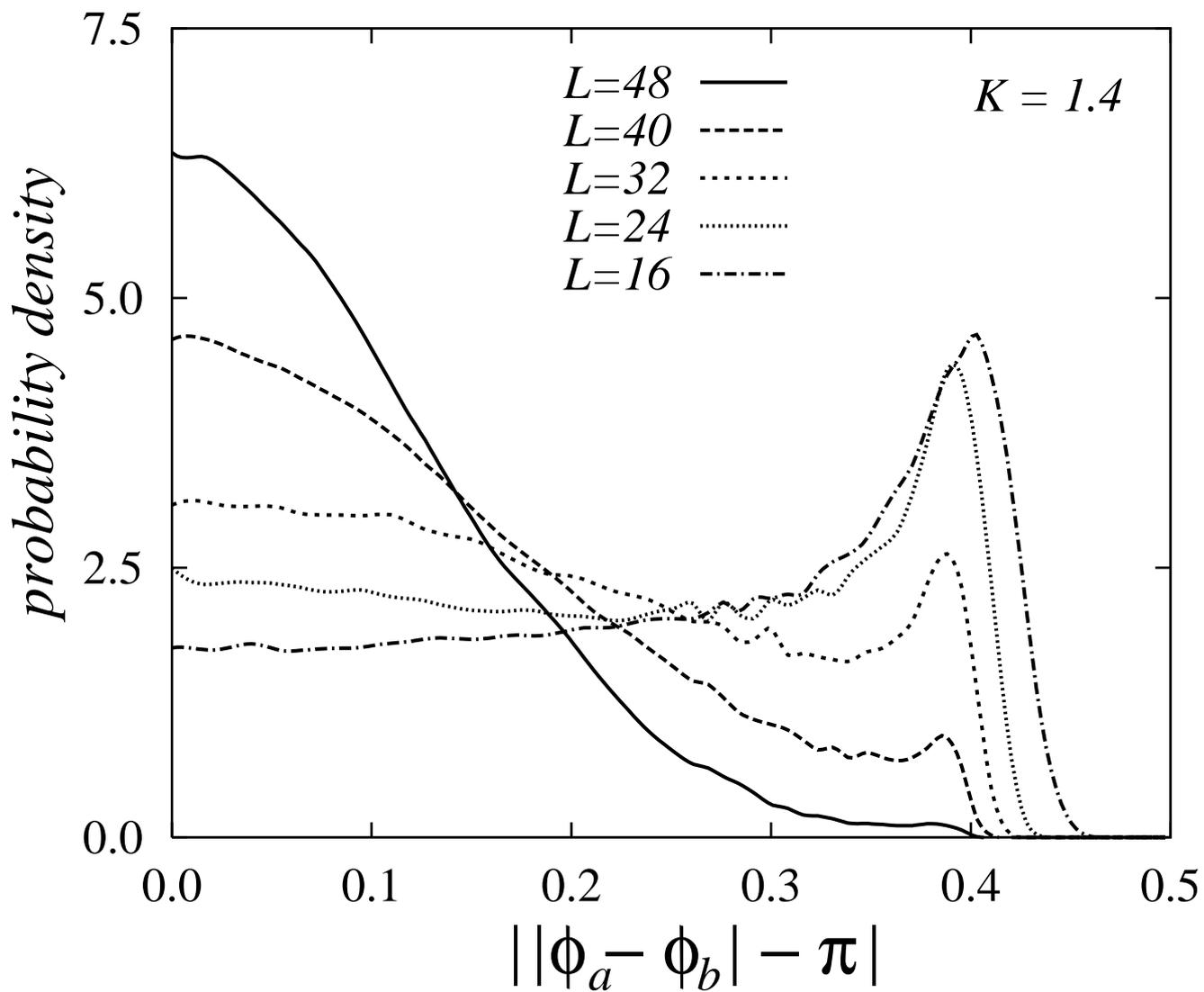

Fig. 5

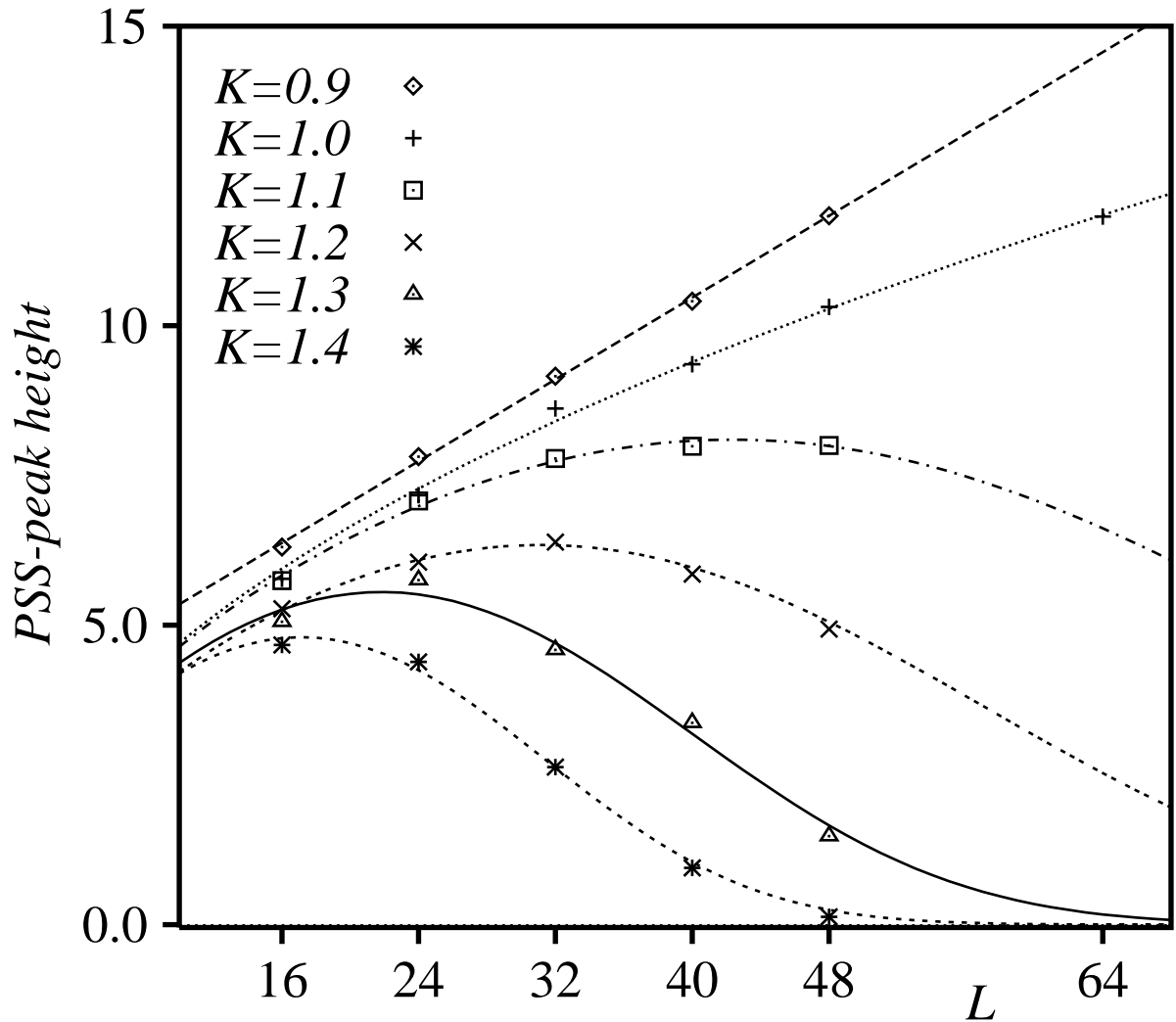

**Fig. 5.** The lattice size dependence of the PSS-peak height in the probability distribution of the $||\phi_a - \phi_b| - \pi|$. The lines correspond to the fits obtained from (7) by adjusting $\delta(T)$ and $v$. The exponent $v$ appears to be near 0.5 for $K \geq 1.0$, and $\delta$ rapidly decreases as $K \to 1.0$. The data are compatible with $\delta = 0$ for $K \leq 1.0$.

Fig. 6

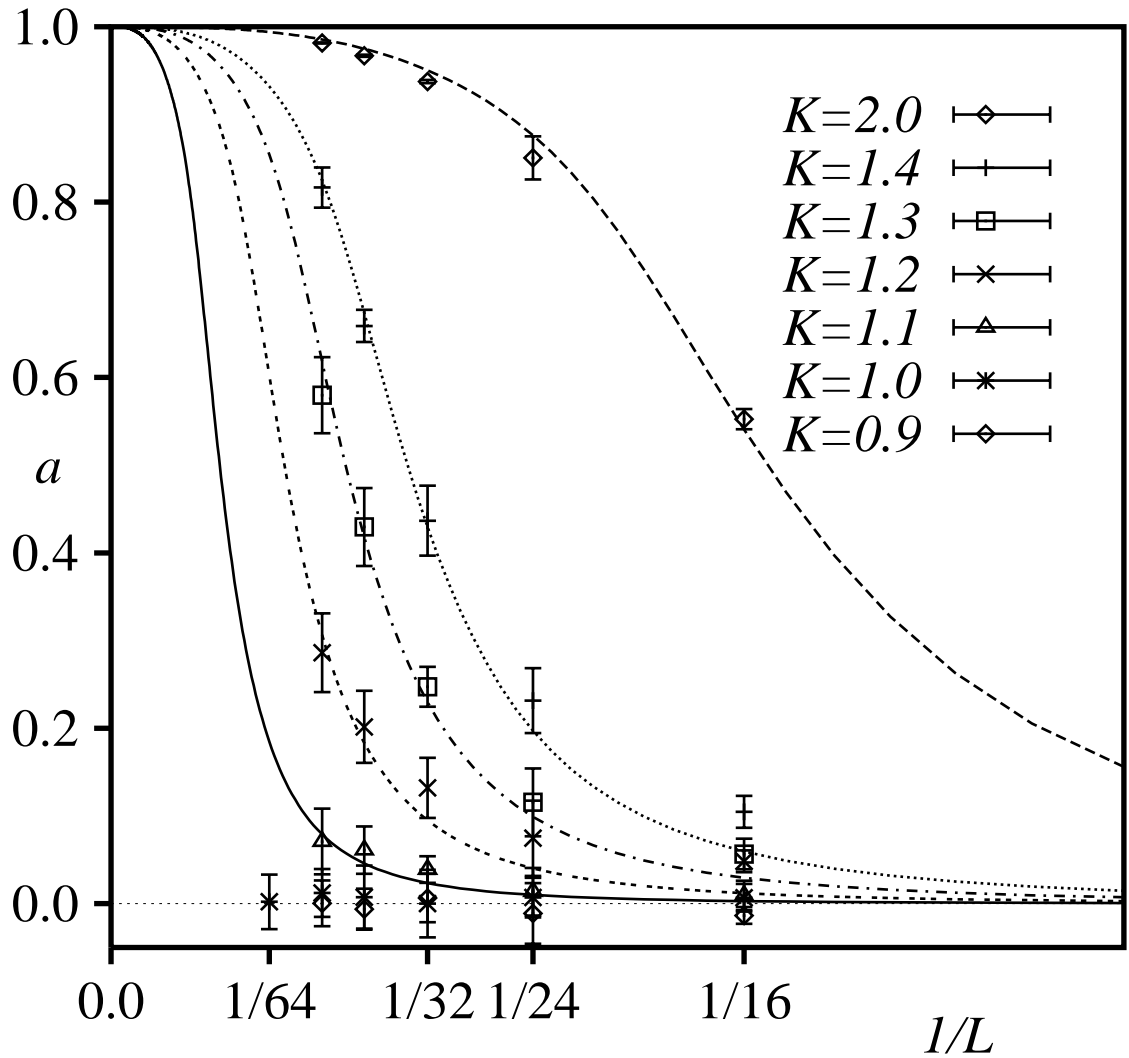

**Fig. 6.** The lattice size dependence of the asymmetry parameter $a$ for several inverse temperatures. The measured data were divided into 5 (10) bins, and the error bars represent the deviations of the resulting mean values. The lines show the fits obtained from the finite-size scaling formula (11).

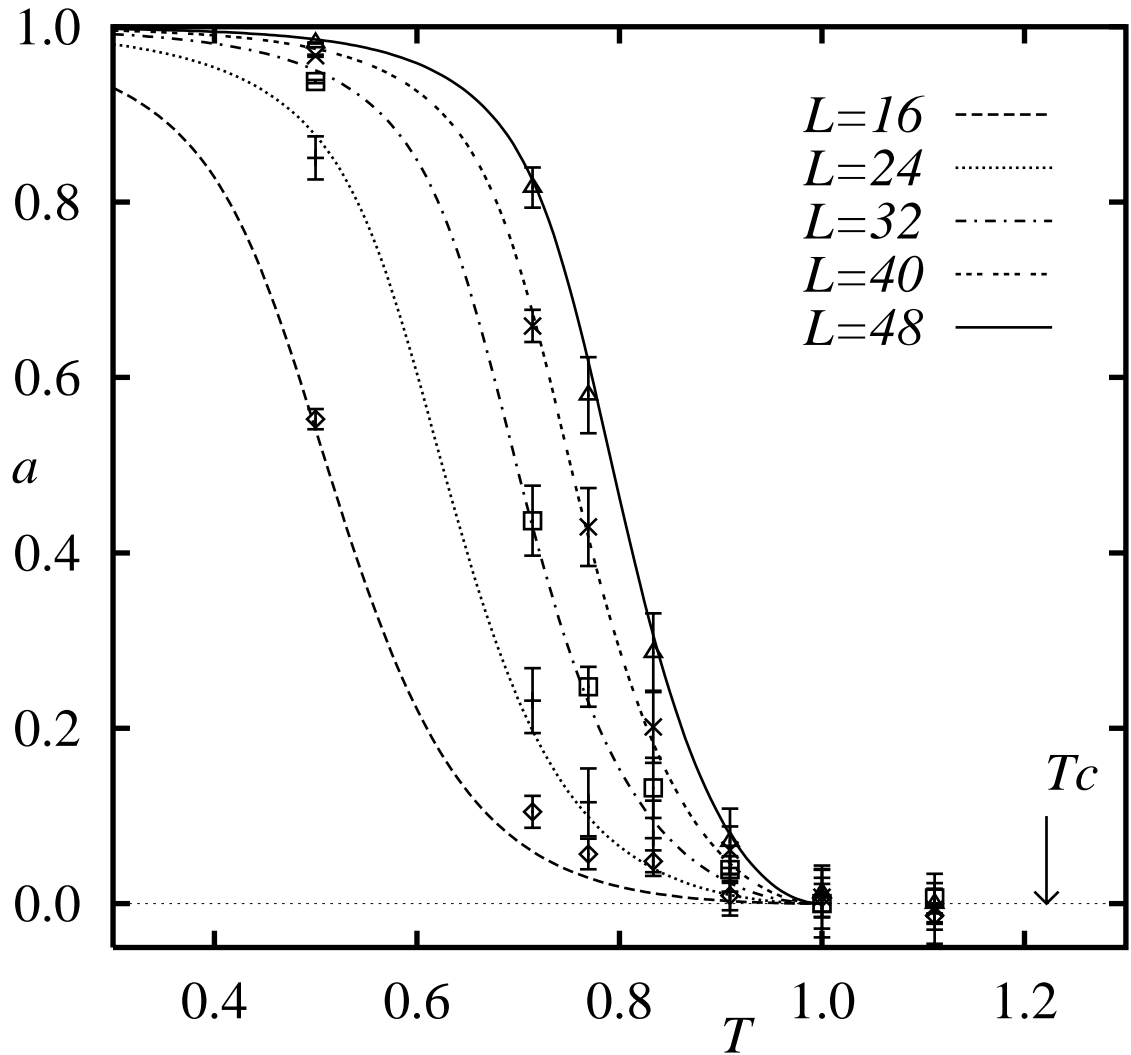

Fig. 7. The parameter $a$ as a function of the temperature. The lines are merely a guide for eyes, and were calculated using the finite-size scaling formula (11) with the free-energy difference $\Delta$ approximated by a polynomial interpolation of the values measured at simulation temperatures. It is seen from this picture that the transition temperature between the BSS- and the rotationally symmetric phases is probably close to $T = 1.0$. The critical point $T_c$ between the ordered and disordered phases is also shown.

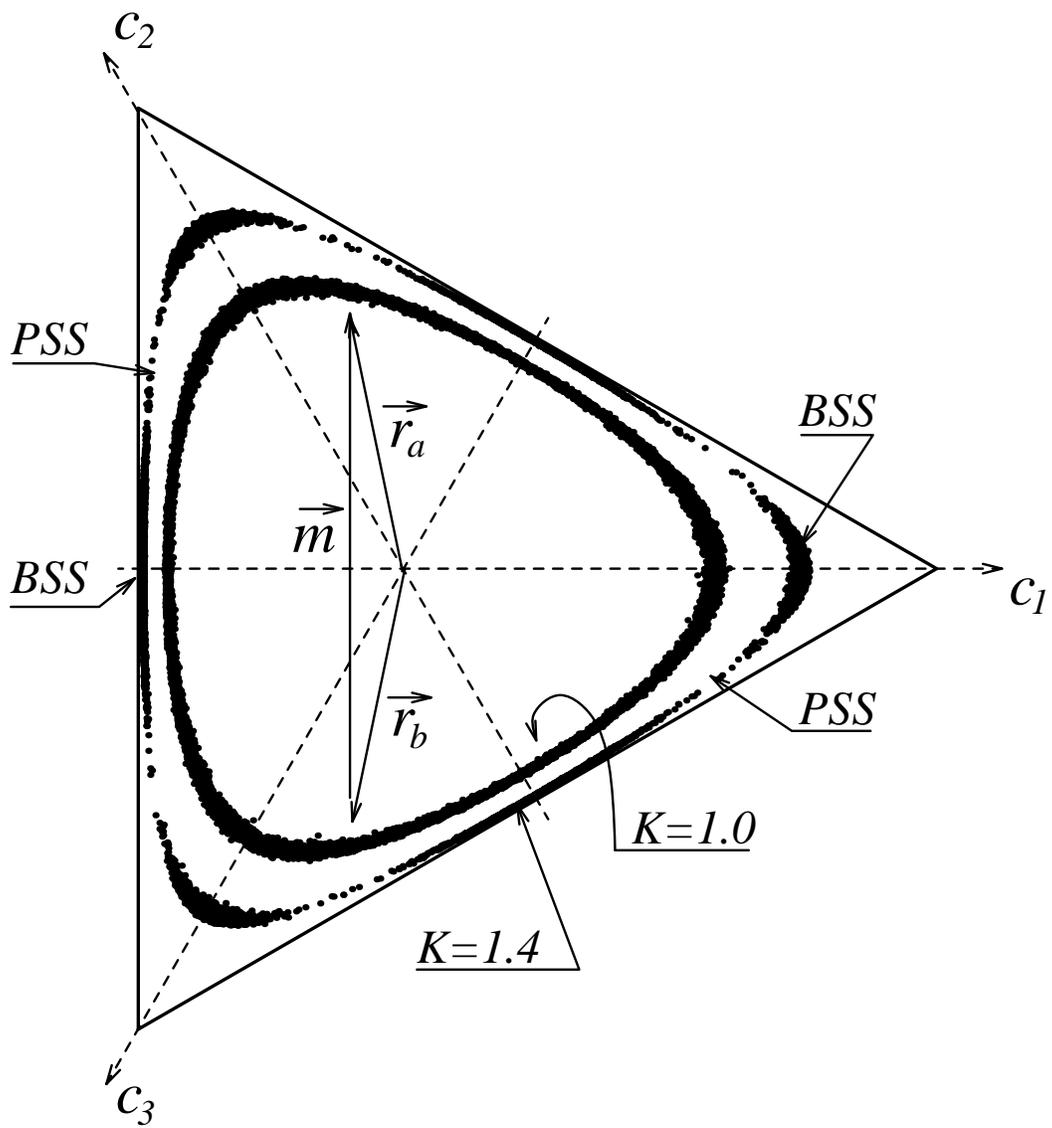

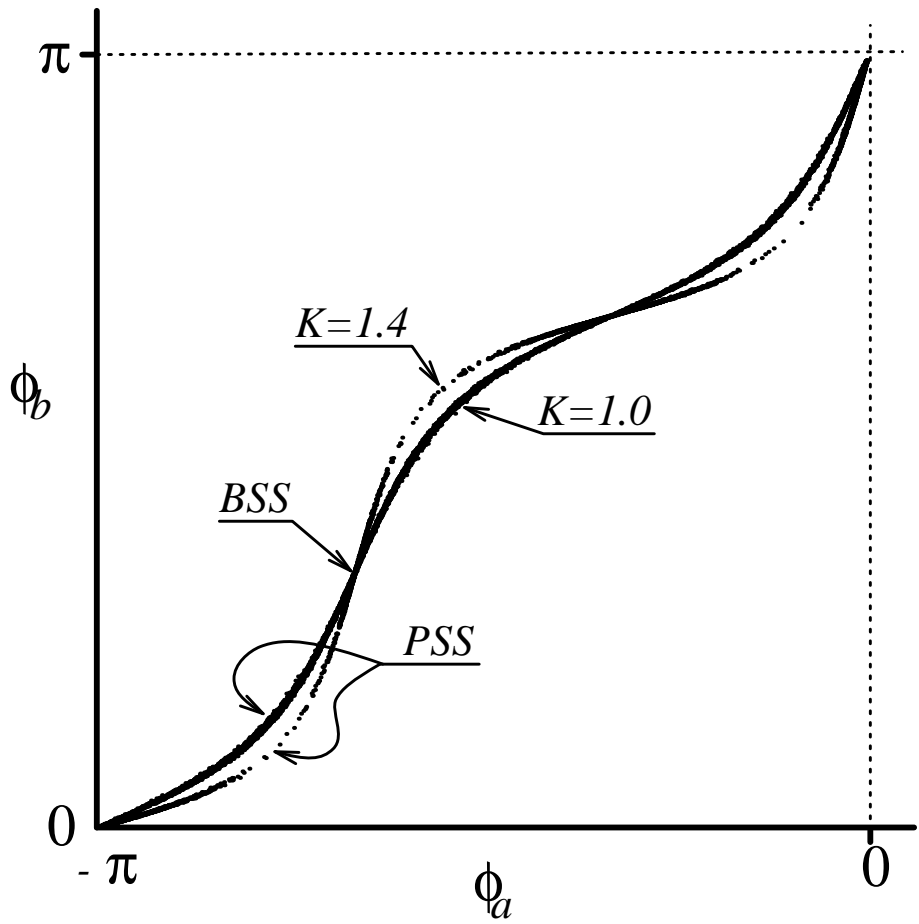